\documentclass{PoS}

\title{Emission line - radio correlation for Low Luminosity Compact sources. 
Evolution schemes.}

\ShortTitle{Emission line - radio correlation for LLC sources}

\author{\speaker{M. Kunert-Bajraszewska}\\
Toru\'n Centre for Astronomy, N. Copernicus University,
Gagarina 11, 87-100 Toru\'n, Poland\\
E-mail: \email{magda@astro.uni.torun.pl}}

\author{Alvaro Labiano\\
ESA, European Space Astronomy Centre (ESAC), 28691 Villanueva de la
Canada, Madrid, Spain}
\author{Marcin Gawro\'nski\\
Toru\'n Centre for Astronomy, N. Copernicus University,       
Gagarina 11, 87-100 Toru\'n, Poland}

\abstract{
We present a radio and optical analysis of a sample of Low Luminosity
Compact (LLC) objects, selected from FIRST survey and observed with MERLIN
at L-band and C-band. The main criterion used for selection was luminosity
of the objects and approximately one third of the CSS sources from the new
sample have a value of radio luminosity comparable to FR\,Is.The analysis of a
radio properties of LLC sources show they occupy the space in radio power
versus linear size diagram below the main evolutionary path of radio
objects. We suggest that many of them might be short-lived objects, and
their radio emission may be disrupted several times before becoming FR\,IIs.
The optical analysis of the LLC sources were made based on the available
SDSS images and spectra. We have classified the sources as high and low
excitation galaxies (HEG and LEG, respectively). The optical and radio
properties of the LLC sample are in general consistent with brighter CSSs
and large-scale radio sources. However, when LLC are added to the other
samples, HEG and LEG seem to follow independent, parallel evolutionary
tracks. LLC and luminous CSS behave like FR\,II sources, while FR\,I seem to
belong to a different group of objects, concerning ionization mechanisms.
Based on our results, we propose the independent, parallel evolutionary
tracks for HEG and LEG sources, evolving from GPS - CSS - FR.
}

\FullConference{10th European VLBI Network Symposium and EVN Users Meeting: VLBI and the 
new generation of radio arrays\\
September 20-24, 2010\\
Manchester Uk}

\begin{document}

\section{Introduction}
Radio sources are divided into two distinct morphological groups of objects:
FR\,Is and FR\,IIs (Fanaroff \&\-Riley, 1974). There is a relatively sharp luminosity
boundary between them at low frequency. The nature of the FR-division is
still an open
issue, as are the details of the evolutionary process in which  
younger and smaller Gigahertz-Peaked Spectrum (GPS) and Compact Steep
Spectrum (CSS) sources become large scale radio structures. The GPS and CSS
sources form a well-defined class of compact radio objects
and are considered to be entirely contained within the host galaxy. During
their evolution the radio jets start to cross the ISM and try to leave
the host galaxy. The interaction with the ISM can be very strong in GPS/CSS
sources 
and it seems to be a crucial point in the evolution of radio sources.

In this paper we present a short summary of the obtained results of the
radio observations of 44 low luminosity compact (LLC) sources and analysis of their optical
properties. The details of the selection criteria and process are given in
Kunert-Bajraszewska et al.(2010a), as are the detailed results of the study of their radio properties.
The details of the optical analysis of the LLC sources are given in
Kunert-Bajraszewska et al.(2010b).
The radio observations of the whole sample were made
with MERLIN at L-band and C-band. Optical data are available for 29 LLC sources and
have been obtained from Sloan Digital Sky Survey (SDSS)/DR7.

\begin{figure*}[t]
\centering
\includegraphics[width=6cm, height=5cm]{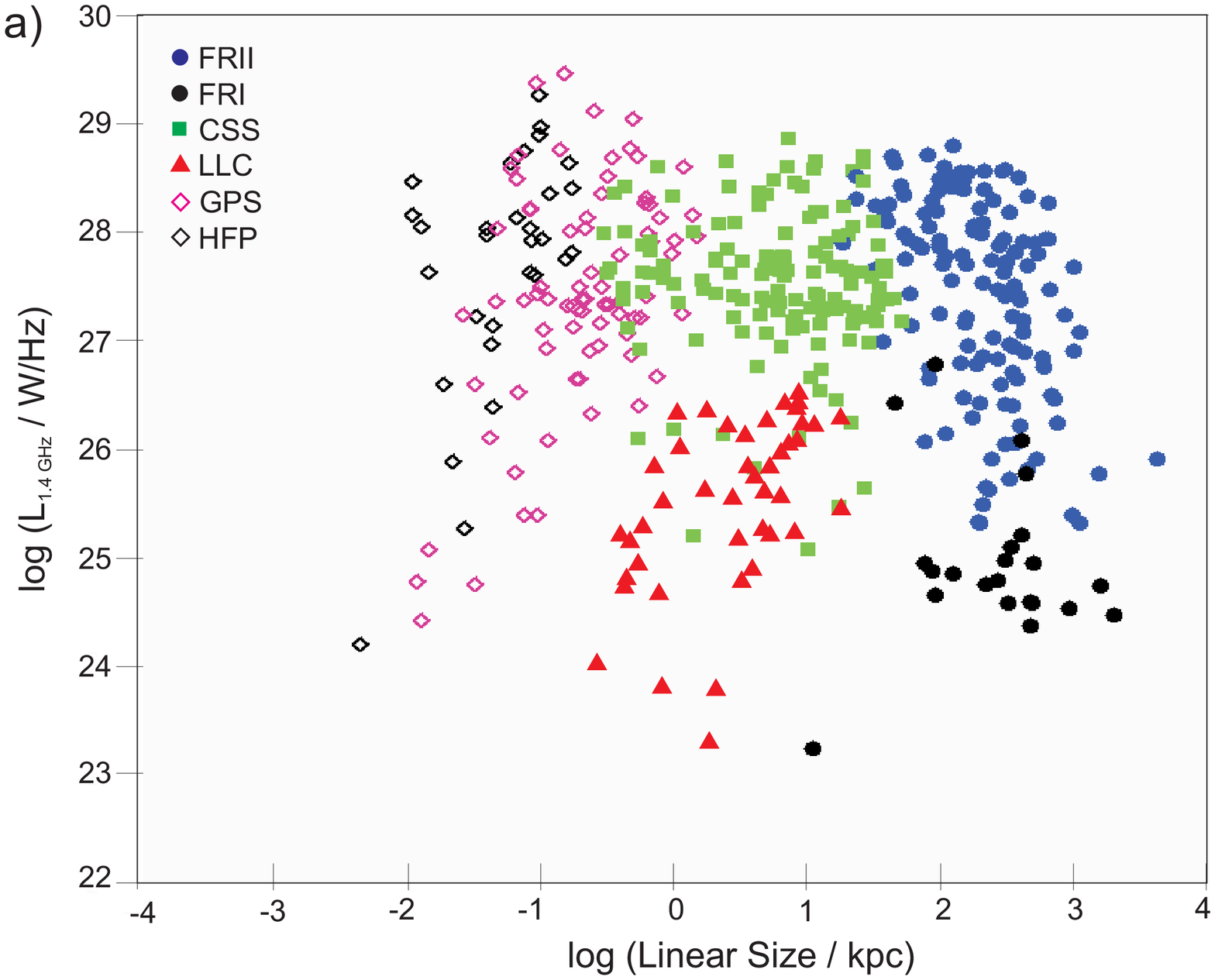}   
\includegraphics[width=7.5cm, height=5cm]{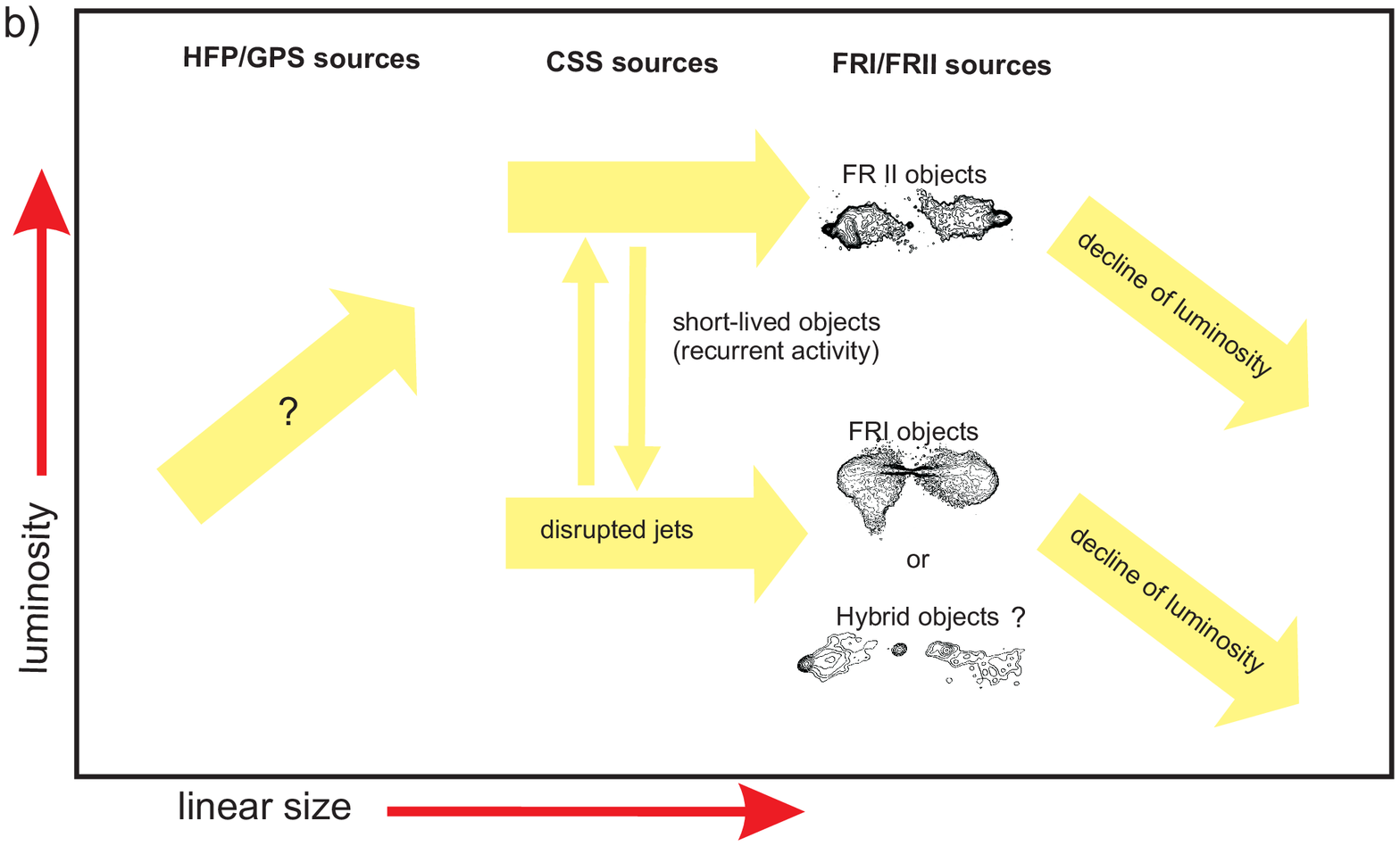}
\caption{(a) Luminosity-size diagram for AGNs. Squares indicate CSS sources
from samples of Labiano et al.(2007), Laing et al.(1983), Willott et 
al.(1999), Fanti et al.(2001) and Marecki et al.(2003).  The diamonds
indicate GPS and HFP objects from sample Labiano et al.(2007). The circles
indicate FRI and FRII objects from samples of Laing et al.(1983),
Buttiglione et al.(2009a,b, 2010)  and Willot et al.(1999).  The triangles
indicate the current sample of LLC sources. (b) Evolutionary scheme of
radio-loud AGNs.}
\label{radio}
\end{figure*}

\section{Results of radio observations}
About 70\% of the observed LLC sources are galaxies and all of them are
nearby objects with redshifts in the range 0.04<z<0.9. Most of them have
been resolved and about 30\% of them have weak extended emission and
disturbed structures when compared with the observations of higher
luminosity CSS sources.We suggest that some of the sources with the breaking
up structures or one-sided morphology are candidates for compact faders. 
We studied correlation between radio power and linear size, and redshift
with a larger sample that included also published samples of compact objects
and large scale FR\,IIs and FR\,Is (Kunert-Bajraszewska et al.,2010a). 
The Luminosity-Size diagram (Fig.1a) shows an evolutionary
scheme of radio-loud AGNs.The selection criteria used for the new sample
resulted in approximately one third of the LLC sources having a value of the
1.4 GHz radio luminosity comparable to FR\,Is. Their luminosities are
definitely lower than CSS sources from last existing samples (Fanti et
al.2001 and Marecki et al.2003). We conclude that many of them can be
short-lived objects, at least in the current phase of evolution and undergo
disrupted evolution many times as they will be able to get out of the host
galaxy and evolve to FR\,IIs (two-sided arrows in Fig.1b). 
The observed parameters of LLC sources (radio luminosity and size), their
radio morphology and spectroscopic features indicate that most of them will
evolve finally to FR\,II. However, we suggest that there exists a much larger
population of short-lived low luminosity compact objects unexplored so far
and among them we can find precursors of large scale FR\,Is. The proposed
evolutionary scheme is drawn in Fig.1b.

\begin{figure*}[t]
\centering
\includegraphics[width=13.5cm, height=6cm]{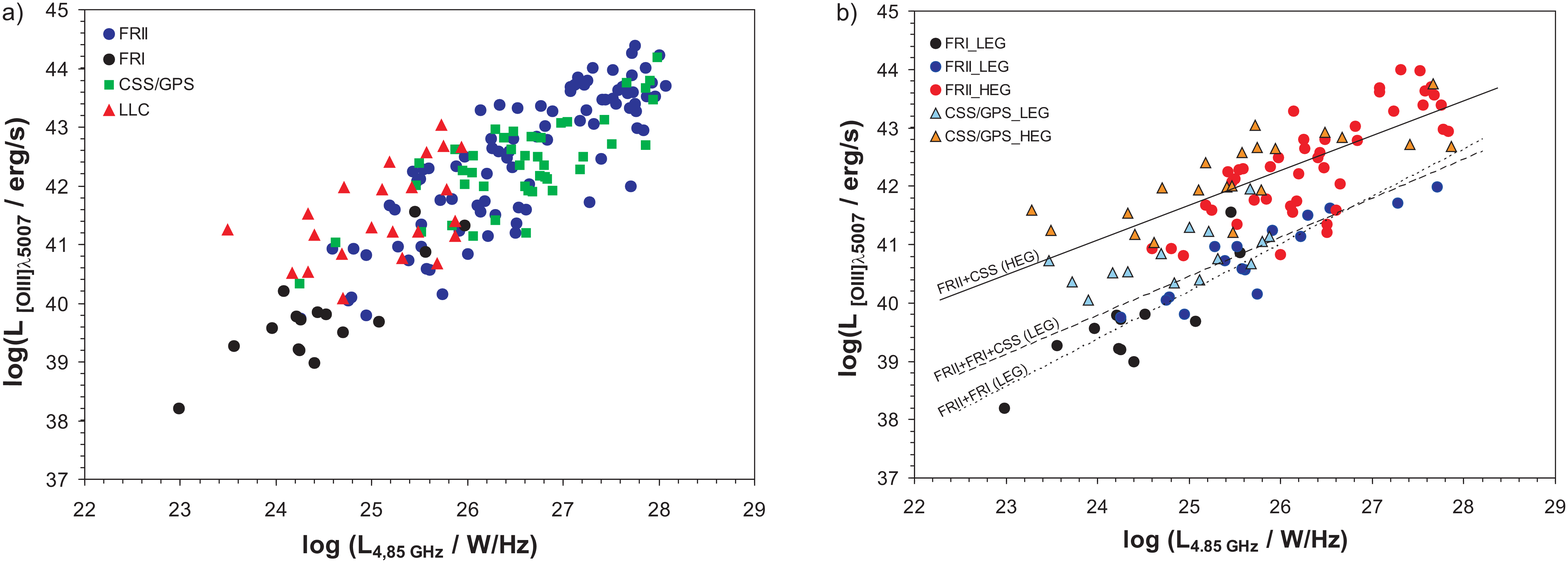}
\caption{(a) [OIII] luminosity - radio luminosity diagram for AGNs, (b)
[OIII] luminosity - radio luminosity diagram for AGNs classified as HEG and
LEG. Squares indicate CSS sources from samples of Labiano et al.(2007),
Laing et al.(1983), Willott et al.(1999), Fanti et al.(2001) and Marecki et
al.(2003).  The diamonds indicate GPS and HFP objects from sample Labiano et
al.(2007). The circles indicate FRI and FRII objects from samples of Laing  
et al.(1983), Buttiglione et al.(2009a,b, 2010)  and Willot et al.(1999).  
The triangles indicate the current sample of LLC sources.
}  
\label{optical}
\end{figure*}

\section{Results of optical analysis}
Optical data are available for most of the LLC sources and based on them we
have classified the sources as high and low
excitation galaxies (HEG and LEG, respectively). 
We have compared the [O\,III] luminosity with the radio
properties for LLC sources, and expanded the sample with other CSS, GPS
sources and FR\,I and FR\,II objects (Kunert-Bajraszewska                    
et al., 2010b). The whole sample shows that, for a given
size or radio luminosity, HEG sources are brighter than LEG in the [O\,III]
line by a factor of 10 (Fig.2b). The LLC objects follow the same correlation between
[O\,III] luminosity and radio power, as the rest of the sample, although the
LLC objects have lower values of [O\,III] luminosity than the more powerful
CSS sources (Fig.2a).
Based on the analysis above, we propose a scenario where the differences in
the nature of LEG and HEG (accretion mode or black hole spin) are already
visible in the CSS phase of AGN evolution and determine the evolution of the
source: i.e.  $\rm CSS_{LEG}$  evolve to $\rm FR_{LEG}$, $\rm CSS_{HEG}$ evolve to
$\rm FR_{HEG}$. 
The main
evolution scenario (GPS-CSS-FR\,II) for radio-loud AGNs was proposed years ago. However, once the
HEG/LEG division is included, these sources seem to evolve in parallel:
$\rm GPS_{LEG}$-$\rm CSS_{LEG}$-$\rm FR_{LEG}$ and $\rm GPS_{HEG}$-$\rm CSS_{HEG}$-$\rm FR_{HEG}$. 
Concerning LEG, it is still not clear if $\rm CSS_{LEG}$ would evolve directly to
$\rm FR\,I_{LEG}$ or go through a $\rm FR\,II_{LEG}$ phase before the
$\rm FR\,I_{LEG}$.  As discussed in
Kunert-Bajraszewska et al., 2010a,  there should also exist a group of
short-lived CSS objects with lower radio luminosities. These short-lived
CSSs could probably show the low [O\,III] luminosities seen in FR\,Is.

\bigskip
\noindent
{\footnotesize
{\bf Acknowledgement}\\
\noindent
MERLIN is a UK National Facility operated by
the University of Manchester on behalf of STFC.\\
This work makes use of EURO-VO software, tools or services. The EURO-VO has
been funded by
the European Commission through contract numbers RI031675 (DCA) and 011892
(VO-TECH) under
the 6th Framework Programme and contract number 212104 (AIDA) under the 7th
Framework Programme.)}


\begin{thebibliography}{99}

\bibitem{butti10} Buttiglione S., Capetti, A.,
Celotti, et al.,
\emph{An optical spectroscopic survey of the 3CR sample of radio galaxies
with $z < 0.3$ . II. Spectroscopic classes and accretion modes in radio-loud
AGN}, \ 2010, A\&A, 509, 6

\bibitem{buti} Buttiglione S., Capetti, A.,
Celotti, et al.,\emph{An optical spectroscopic survey of the 3CR sample of
radio galaxies with $z < 0.3$. I. Presentation of the data}, \ 2009b,
A\&A, 495, 1033

\bibitem{buti09} Buttiglione, S.; Celotti, A.;
Capetti, A., et al.,\emph{Optical spectra of 15 Low Luminosity Compact
Sources and the formation of jets}, \ 2009a, AN, 330, 237  

\bibitem{fr74} Fanaroff, B.~L., \& Riley,
J.~M., \emph{The morphology of extragalactic radio sources of high and low
luminosity}, \ 1974, MNRAS 167, 31

\bibitem{f2001} Fanti, C., Pozzi, F., \& Dallacasa, D.,
et~al.,\emph{Multi-frequency VLA observations of a new sample of CSS/GPS
radio sources}, \ 2001, A\&A, 369, 380

\bibitem{kunert10a}
Kunert-Bajraszewska, M., Gawro\'nski, M.~P., Labiano, A., Siemiginowska, A.,
\emph{A survey of low-luminosity compact sources and its implication for the
evolution of radio-loud active galactic nuclei - I. Radio data}, \ 2010a, MNRAS,
[arXiv:1009.5235] 

\bibitem{kunert10b}
Kunert-Bajraszewska, Labiano, A., \emph{A survey of low-luminosity compact
sources and its implication for the 
evolution of radio-loud active galactic nuclei - I. Optical analysis}, \ 2010b, MNRAS,
[arXiv:1009.5237]

\bibitem{labiano07} Labiano, A., Barthel, P.D., O'Dea,
C.P., et al.,\emph{GPS radio sources: new optical observations and an
updated master list}, \ 2007, A\&A, 463, 97

\bibitem{laing83} Laing, R.~A., Riley, J.~M., \&
Longair, M.~S., \emph{Bright radio sources at 178 MHz - Flux densities,
optical identifications and the cosmological evolution of powerful radio
galaxies}, \ 1983, MNRAS 204, 151

\bibitem{mar03} Marecki, A., Niezgoda, J.,
W{\l}odarczak, et al.,\emph{Weak CSS Sources from the FIRST Survey}, \ 2003, PASA 20,
42

\bibitem{willott} Willott, C. J., Rawlings, S.,
Blundell,
K. M., \& Lacy, M., \emph{The emission line-radio correlation for radio
sources using the 7C Redshift Survey}, \ 1999, MNRAS, 309, 1017


\end{thebibliography}
\end{document}